\documentclass[12pt,preprint]{aastex}

\newcommand{\GA}{\mbox{\raisebox{-0.6ex}{$\stackrel{\textstyle>}{\sim}$}}}
\newcommand{\cxo}{{\sl Chandra}}
\newcommand{\ngc}{{NGC~7331}}
\newcommand{\msun}{M$_{\odot}$}
\newcommand{\ergl}{ergs~s$^{-1}$}

\newcommand{\cxou}{CXOU~J223706}

\newcommand{\ha}{H$\alpha$}

\newcommand{\oiii}{[O~{\sc iii}]}
\newcommand{\oi}{[O~{\sc i}]}
\newcommand{\sii}{[S~{\sc ii}]}

\newcommand{\hii}{H~{\sc ii}}
\newcommand{\nii}{N~{\sc ii}}

\newcommand{\hst}{{\sl Hubble}}

\newcommand{\etal}{et al.}

\renewcommand{\mag}[1]{^{\rm m}\!\!\!#1\,}
\newcommand{\AAA}{ \AA \/}

\slugcomment{Submitted to Astrophysical Journal}

\begin{document}

\title{Optical Spectroscopy of the environment of a ULX in NGC 7331}

\author{
Pavel K. Abolmasov\altaffilmark{1}, 
Douglas A. Swartz\altaffilmark{2},
S. Fabrika\altaffilmark{1,3},
Kajal K. Ghosh\altaffilmark{2},
O. Sholukhova\altaffilmark{1}, and
Allyn~F.~Tennant\altaffilmark{4}
}
\altaffiltext{1}{Special Astrophysical Observatory, Nizhnij Arkhyz 
369167, Russia}
\altaffiltext{2}{Universities Space Research Association, NASA Marshall Space Flight Center, VP62, Huntsville, AL 35805, USA}
\altaffiltext{3}{University of Oulu, P.O. Box 3000, 90014, Finland}
\altaffiltext{4}{Space Science Department,
    NASA Marshall Space Flight Center, VP62, Huntsville, AL, USA}

\begin{abstract}

Optical photometric and spectroscopic data are presented that
 show an association of an ultraluminous X-ray source
 in \ngc\ with a young star cluster of
 mass $M = (1.1\pm0.2)\times 10^5$~\msun\ and age
 $t_c=4.25\pm0.25$ Myr.
If the ULX is part of the bright stellar cluster, 
 then the mass of the progenitor of the compact accretor must have been $\gtrsim$40-50~\msun\
 in order to already have evolved through the supernova stage 
 to a compact object. 
The companion star is also likely an evolved massive star.
The emission line spectrum of the nebula surrounding the cluster can be 
 interpreted as a result of photoionization
 by the cluster OB stars with an additional source of shock excitation
 producing strong \sii, \oi\ and \nii\ lines. 
This additional source appears to be as much as five times more powerful than the supernovae and
stellar winds in the cluster can provide.
Additional mechanical energy input associated with the ULX itself 
can help explain the residual shock excited
line luminosities of the emission region. 

\end{abstract}


\section{Introduction}

Ultraluminous X-ray sources (ULXs) are point-like non-nuclear sources in external galaxies
 with apparent X-ray luminosities greater than the Eddington limit of stellar mass
 black holes, i.e., $L_{\rm X}$\GA$10^{39}$~\ergl.
The existence of ULXs has been known since 
 the first {\sl Einstein} observations of nearby galaxies
 \citep{lvs83, helfand84,fabbiano89}
 but only with the sub-arcsecond 
  imaging capability of the \cxo\ X-ray Observatory 
 have investigations at other wavelengths of the crowded 
 fields of nearby galaxies become viable.

Multi-wavelength investigations enable a much broader 
 range of diagnostic techniques in the study of ULXs than can
 be achieved by X-ray data analysis alone.
In particular, traditional methods in 
 optical imaging and spectroscopy can be used to constrain 
 the age, metallicity and physical state of the local environments of ULXs.
This, in turn, can help to determine the origin and history of the ULXs themselves
 and lead to a better understanding of the influence of ULXs on their surroundings.

Many of the models for ULXs favor young high-mass X-ray binary systems
 (see, e.g., the reviews by \citet{fw2006}; \citet{fabbiano2006}).
These objects should be preferentially associated with
 massive star forming regions such as OB associations and,
 at the high-mass limit, superstar clusters.
These environments should be strong sources of line emission
 through photoionization by massive stars, supernovae (SNe) and by the ULX itself
 and may display enhanced forbidden line emission from
 shock excitation by stellar winds, SNe, and by energetic outflows and jets that
 could accompany the ULX X-ray emission.
Nebulae such as those associated with the ULXs in Ho~IX 
 \citep{GPM1},
 NGC 5408 \citep{soria5408}, NGC 5204 \citep{goad5402},  and Ho~II
 \citep{hoii_kaaret,Leh_ea05} are among the best-studied
 examples of this intricate ULX/star-formation connection.

As part of a program of spectroscopic studies of the 
 local envrironments of ULX candidates, we have obtained a moderate-resolution
 long-slit spectrum of a ULX in \ngc.
The X-ray characteristics of this ULX and the optical photometric properties of 
 its environment are described in \S~\ref{s:OBS}. 
The ULX lies roughly in the center of an extended \hii\ region,
cataloged as P98 \citep{Petit98},   
 and within 
 30~pc of the center of a bright blue emission region. 
The optical spectrum of this emission region is presented 
 in \S~\ref{s:SPEC}.
We model its spectrum as emission from a young star cluster combined with a
nebula photoionized by a cluster hosting an X-ray source (\S~\ref{s:RESULTS}).
This model reproduces the observed $4000 - 7300$~\AAA\ continuum spectrum but 
 under-predicts several strong observed emission lines. 
We argue (\S~\ref{s:shocks}) these lines are due, at least in part, to shock excitation
 associated with the ULX and that radiative and mechanical feedback plays
 an important part in the overall energetics of the emission region.
We briefly summarize our results and compare the 
 properties of the environment of the ULX in \ngc\ to other ULX nebulae
 in \S~\ref{s:discussion}.

\section{X-ray and Optical Observations} \label{s:OBS}

\ngc\ is a highly-inclined, i = 77$^\circ$ \citep{ggal},
SA(s)b spiral galaxy subtending 10.\arcmin 5
on the sky. We adopt a distance of 14.6~Mpc \citep{ferrarese};
1\arcsec\ corresponds to 71~pc.

\begin{figure}
\includegraphics[angle=0,width=0.97\columnwidth]{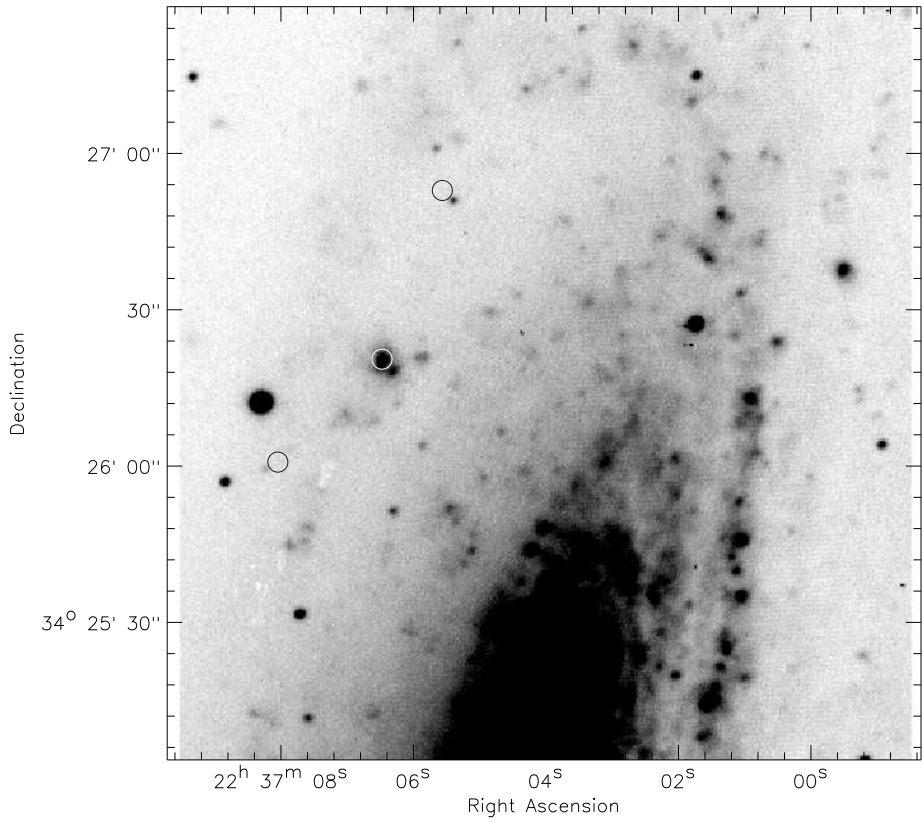}
\figcaption{Isaac Newton Telescope 
 \ha\ image of a portion of the \ngc\ field showing the location of 
 three ULXs in relation to the central regions of the galaxy. 
At the distance of 14.6~Mpc to \ngc, 1\arcmin\ corresponds to 4.25~kpc.
Circles denote  2\arcsec\ radius locations
 of the three ULX candidates detected. \cxou\ is the middle source.
 \label{f:ha_image} } 
\end{figure}

Swartz \etal\ (2004) identified three ULX candidates within the $D_{25}$
 isophote of the \cxo\ field of \ngc. 
All three lie in the northeast portion of the spiral arm structure
 beyond the large ring seen in the Spitzer Infrared Nearby Galaxies Survey 
 images  
 \citep{sings}.
An \ha\ image of this region is shown in Figure~\ref{f:ha_image} with the
 approximate positions of the three ULX candidates superposed.
Only the brightest ULX candidate has an obvious association with optical
 emission features in this or any other images we have studied. 
For the remainder of this work, we will refer to this ULX, which is formally
 designated CXOU~J223706.6+342620, as \cxou.

\subsection{\cxo\ observations}

\cxo\ obtained a 29.5~ks observation of a portion of \ngc\ on 12 Aug 2001
(ObsID 2198).
The observation was taken in sub-array mode using the Advanced CCD Imaging 
 Spectrometer (ACIS) with the galaxy center imaged near the aimpoint on the 
 back-illuminated S3 CCD.
\cxou\ is approximately 1.\arcmin 85 off-axis in this observation.
Using the known position of the nucleus\footnote{$\alpha = 22^h37^m04^s\!.096, \ \delta = +34^\circ24^\prime56^{\prime\prime}\!.29$ (J2000); \citet{opos}.}  of \ngc, 
we estimate the absolute positional
 uncertainty of the ULX is $\sim0\farcs4$.
Within a 3$\sigma = 1\farcs86$ circle about the position of the source,
 175 X-ray events were detected corresponding to a S/N of 10.6 
 (in the 0.5--8.0 keV band).
Acceptable fits to the spectrum can be made using a variety of models. The
 best-fitting model attempted is an absorbed power law 
 ($N_{H} = 6.9_{-2.4}^{+3.1} \times 10^{21}$~cm$^{-2}$, $\Gamma = 2.26_{-0.52}^{+0.63}$,
$\chi^2=5.1$ for 5 dof after grouping to ensure at least 20~counts per spectral bin).
The corresponding intrinsic source luminosity is 
 $L_{\rm X} = (2.8\pm0.6)\times 10^{39}$~\ergl\ 
 according to our previous analysis \citep{swartz}.  
(Here and elsewhere, errors are extremes on the single interesting parameter 1$\sigma$ confidence intervals.)
The luminosity of \cxou\ lies at the median value of the 97 ULX candidates in our sample 
 of ULXs in spiral galaxies. 
The average luminosity for that subsample is $5.9\times 10^{39}$~\ergl\
 and the highest luminosities exceeded several $10^{40}$~\ergl.
If we assume the X-ray luminosity approximates the bolometric luminosity and 
 that the source radiates isotropically at the Eddington limit
 during the \cxo\ observation, then the 
 mass of the compact accretor is 22$\pm$5~\msun.

\subsection{\hst\ observations}\label{s:hst}

\hst/WFPC2 observations\footnote{Available from http://archive.stsci.edu/} of NGC7331 were carried out on 
 1997 August 13 and 14 using F450W and F814W filters.
Astrometric corrections were made using USNO stars
\citep{usno}. 
The 1$\sigma$ nominal
absolute positional uncertainities in the \hst\ images are 
$0\farcs3$ giving a combined uncertainty of 
about $0\farcs5$ between the \hst\ and \cxo\ registration.

\begin{figure}
\includegraphics[width=1.0\columnwidth]{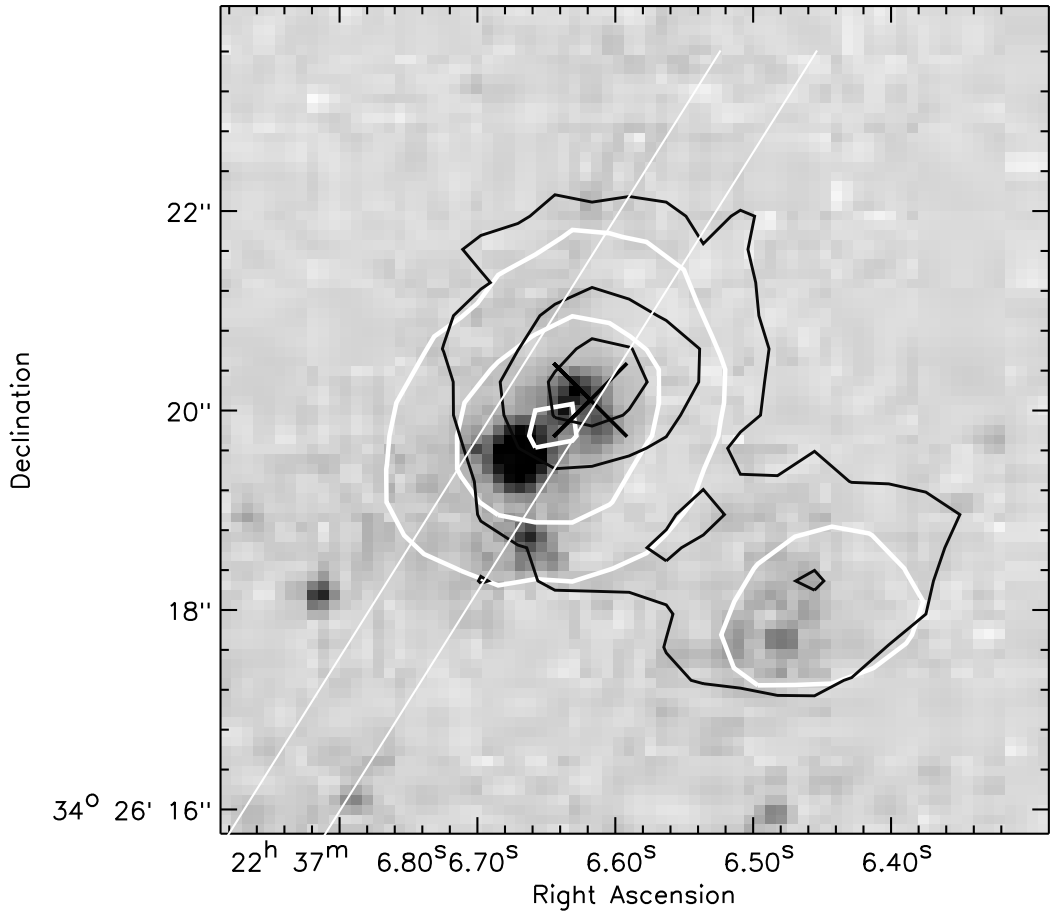}
\figcaption{
8\arcsec $\times$8\arcsec\ \hst\ F450W image of the field around the ULX source \cxou. 
The X-ray source position is marked by the "X" of $\pm 0\farcs5$ radius 
 corresponding to the combined \cxo\ and \hst\ positional uncertainty. 
Ground-based \ha\  (black) and \oiii\ $\lambda 5007$ (white)
isophotes at the 20, 50 and 85\% levels above their backgrounds 
are superimposed (see \S~\ref{s:INT}). The slit position of the optical 
spectrum (\S~\ref{s:SPEC}) is also shown.
    \label{f:FOV} }
\end{figure}

Figure~\ref{f:FOV} displays an 8\arcsec $\times$8\arcsec\ 
 portion of the F450W \hst\ image around \cxou.  
The \hst\ \ image is dominated by two bright extended sources; presumably young
 stellar clusters. The X-ray source
 is located within the NW cluster, within the formal error.
The optical source has a FWHM of $\sim0\farcs1$ or 7~pc at the distance of \ngc\
(the source is not completely symmetric, \hst\ images reveal 
additional emission in the wing).
The total background-corrected brightness
within a $1\farcs5$ diameter circle (equivalent to the seeing conditions in the 
 ground-based data, see below) centered on the X-ray source position
 is $m_{\rm F450W}=20.44$~mag.
This corresponds to an absolute magnitude $M_{450W} = -10.4$~mag,
 lying at about the maximal luminosity of the Galactic open 
 clusters \citep{larsen}. 
However, taking into account interstellar absorption
 (\S~\ref{s:RESULTS}), the object intrinsic brightness may be significantly higher.

\subsection{Isaac Newton Telescope observations} \label{s:INT}

Archival optical images of \ngc, obtained with the Wide Field Camera (WFC) on the
Isaac Newton Telescope (INT) at La~Palma\footnote{Available from http://casu.ast.cam.ac.uk/casuadc/archives/ingarch}, 
were examined for structure local to \cxou.
We used images taken in narrow filters \oiii\ (central wavelength 5008~\AA, FWHM=100~\AA),
H$\alpha$ (6568~\AA, 95~\AA) and in the Harris broad-band filters B (4298~\AA, 1065~\AA),
V (5425~\AA, 975~\AA) and R (6380~\AA, 1520~\AA). All the imaging observations were
carried out in 2000 and 2001. We have estimated seeing from the FWHM of unsaturated
 point-like sources in the images to be 
$0\farcs7$ in H$\alpha$, $0\farcs8$ in \oiii\ and from $0\farcs9$ to $1\farcs2$ in the broad bands. 

Astrometric corrections were made using USNO-B1 and USNO-B2 standards 
\citep{usno}. The astrometrical error obtained in all these images 
(using 22 to 27 astrometrical standard 
 stars) is less than $0\farcs2$.
We made photometric calibration of the B- and V-band images using 
19 USNO standard stars that range in brightness from 13.0 to 20.5 mag.
A systematic error of the USNO photometric standards as high as 0.2-0.3 
magnitude is possible for the fainter stars \citep{usno}. 

We created new \ha\ and \oiii\ images free of continuum emission by dividing the narrow band 
images over the R and V images, respectively. In the resulting images the majority 
of single stars (presumably non-emission-line stars) disappear.
From these images one may estimate the spatial structure in the emission lines
and measure the corresponding emission line equivalent widths.
Isophotes at the 20, 50 and 85\% levels above the background in these \ha\ and \oiii\ 
images are shown in Figure~\ref{f:FOV} (100\% corresponds to the maximum value in the 
8\arcsec$\times$8\arcsec\ field). 
Taking the 50\% isophote as representative of the extent 
of the emission in these lines, one obtains the size of the emission-line
region to be $\sim$2\arcsec\ in \ha\ and \oiii\ 
corresponding to 140 pc (FWHM).
The peak of the \ha\ emission is coincident with the NW cluster resolved in the \hst\ image.
The peak \oiii\ emission lies between the two clusters suggesting each cluster contributes 
 equally to the \oiii\ emission.  

The estimated B-band magnitude within a circle of $1\farcs5$ diameter around the NW
cluster is $19\mag{.}97\pm0.06$ (USNO-B2) and $20\mag{.}32\pm0.06$ (USNO-B1).
For modelling the optical spectrum (see below),
we adopt the value B=$20\mag{.}0$.

\section{Optical Spectroscopy} \label{s:SPEC}

Optical spectral data were obtained using the 6m Russian telescope 
 with the SCORPIO 
focal reducer \citep{scorpio} in the long-slit (LS) mode  
with a $0\farcs75$-wide slit, providing
spectral resolution of about 10 \AA.
The data cover the spectral range $4000-7300$ \AA.
The object was observed on August 29, 2005
at $1\farcs4$ seeing conditions.
The reduction process includes all the standard procedures.

The spectrum is shown in Figure~\ref{fig:ospect_mod}. 
Many strong emission lines are present above a well-defined blue continuum.
Table~1 lists the observed and dereddened integral fluxes in all the emission
lines detected.

We extracted slit images in the strongest lines (H$\beta$, \oiii, \ha, and \sii).
The extent of the \oiii\ and \ha\ emission in the slit images is consistent 
 with the structure seen in the INT images.
The emission in the slit images can best be described as the sum of a point-like
structure (FWHM$\sim1\farcs5$ consistent with the seeing conditions)
and more extended (FWHM$\sim5\farcs0$) emission. 
That is, the two clusters are not resolved.
The centroids of the point-like emission 
in all the lines are coincident within their errors;
but in the extended component the \oiii\ centroid 
is shifted about $0\farcs4$ to the SSE relative to \ha\ along the slit 
(cf. Figure~\ref{f:FOV}). The \sii\ centroid is shifted in the opposite direction,
about $0\farcs2$ to the NNW relative to the \ha\ peak.

\section{Spectrum modelling} \label{s:RESULTS}

The size and luminosity of the bright sources in the \hst\ blue (F450W) image 
 suggest they are compact young star clusters. 
The INT narrow-band images and SAO slit images suggest the star clusters are 
embedded in a larger emission line nebula
 (with the slit sampling most of the cluster emission but not all of the 
 larger emission line region).
We constructed synthetic spectra based on a model for the emission 
 consisting of an underlying young stellar population 
 and a photoionized nebula. 
Starburst99 \citep{sb2005} was used for the stellar contribution which 
 provides model spectra for the continuum emission as functions of age
 and metallicity of the cluster and normalized by the cluster mass.
CLOUDY96.01 \citep{cloudy} was used to simulate the emission line spectrum
 from the photoionized nebula as a function of the input photoionizing source
 spectrum and ionization parameter, the nebular abundances, and the density.
The spectral modelling began by finding the best-fitting 
 cluster model. This model was then subtracted from the observed spectrum
 and the best-fitting model 
 for the residual spectrum was found using the photoionization code.

\subsection{The contribution from the Star Cluster}

\begin{figure}
\includegraphics[width=\textwidth]{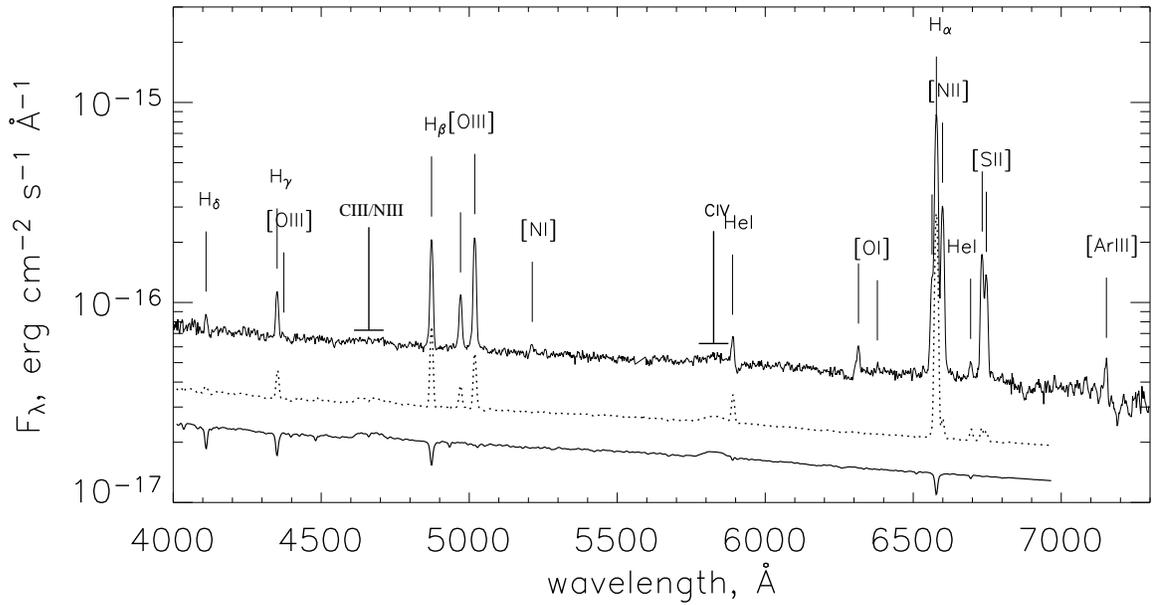}
\figcaption{Integral spectrum of the cluster near \cxou\ (top). 
The lower thick solid curve 
is the best-fitting stellar spectrum model estimate,
divided by a factor of 3 for clarity, and the dotted curve is the combination of this model 
and the best-fitting photoionization model (see \S~\ref{s:RESULTS}),
divided by a factor of 2. The brightest emission lines including
Wolf-Rayet blends are superscribed.
  \label{fig:ospect_mod} }
\end{figure}

The oxygen abundance and abundance gradient in \ngc\ is given by \citet{pilyugin}. 
At the location of \cxou, about 90\arcsec\ from the center
 of the galaxy, 
$12+\log ({\rm O/H}) = 8.32$ or about 2--3 times less than the solar value.
Using this as a guide,
we constructed Starburst99 models with [Fe/H]$=$0.0, $-0.4$, $-0.7$.
We computed a grid of models for cluster ages, $t_c$, ranging from 1 to 30~Myr.
The model spectra were fit to the observed spectra allowing for a 
variable interstellar absorption component (using the reddening curve of 
\citet{CCM}).
Since the Starburst99 stellar cluster model does not predict 
a nebular emission line spectrum it was necessary to exclude 
all the bright observed emission lines in the model fitting. 
We find the best-fit parameters of the cluster model are
 $t_c=(4.25\pm0.25)$~Myr, $A_V = 1\mag{.}43 \pm 0\mag{.}05$,
 $M = (1.1\pm 0.2)\times10^5$~\msun\ where $M$ is the cluster mass 
 which scales linearly with the overall model normalization. 
The $\chi_r^2$ statistic for this model is 1.30 for $\sim$200 degrees of freedom
(the spectral range divided by the spectral resolution element with emission 
lines omitted and reduced by the number of model fit parameters). 
Solar metallicity provides a significantly better fit.
The best-fit model spectrum (scaled by a factor of $1/3$ for clarity) is shown
as the lower curve in Figure~\ref{fig:ospect_mod}.

The age and metallicity derived here are consistent with the presence of
the broad but weak emission features observed at 4650~\AA\ and 5806~\AA.
We interpret these features as the C~{\sc iii}$+$N~{\sc iii} and
 C~{\sc iv} line blends, respectively, due to the presence of Wolf-Rayet 
stars \citep{wr_conti,SV98} in the cluster. We estimate the equivalent
widths to be  1.7$\pm$0.2~\AA\ for the 
blue blend and 4.2$\pm$0.2~\AA\ for the red blend of lines. 
(The quoted errors are formal errors using a piece-wise linear approximation to
the local continua; inspection of Figure~\ref{fig:ospect_mod} suggests 
the uncertainties are likely higher.)
The theoretical blue-to-red ratio \citep{SV98} 
most closely approaches the observed ratio at 3--6 Myr where the 
theoretical ratio is $\approx$1 for a wide range of metallicities.

\subsection{The contribution from photoionization}

The best-fit stellar population model spectrum was subtracted 
 from the dereddened observed spectrum to give an estimate 
 of the purely nebular spectrum. 
We used the $3\le t_c \le 6$~Myr
 UV spectrum from the best-fit unabsorbed cluster model plus an additional
 X-ray component based on the \cxo\ spectrum of \cxou\
 ($\Gamma=2.2$, 3.2$\times$10$^{48}$~photons s$^{-1}$ blueward of 
 the Ly~edge) for the photoionization source for the nebula model.
We calculated a set of spherically-symmetric photoionization
 models with fixed inner (1~pc) and outer (30~pc) radii using CLOUDY96.01. 
The inner radius is typical for a massive young star cluster 
(e.g., \citet{OCnilakshi}).
The outer radius corresponds to the
 half-width of the slit in projection (at the galaxy distance of 14.6~Mpc).
We have found that the choice of
 inner and outer radii do not strongly affect the results.
The nebula gas density and the low-energy cutoff for the ionizing 
 X-ray spectrum were allowed to vary in the modelling.
The best-fitting gas density is about 30~cm$^{-3}$
 (consistent with the observed \sii\ $\lambda6717/\lambda6731$ ratio).
The models proved insensitive to 
 the additional X-ray and corresponding EUV radiation from the ULX.
This was anticipated because the number flux from the cluster model
 in the Ly~continuum 
 ($\sim$10$^{51}$~s$^{-1}$) far exceeds the flux of X-ray photons
 even after extrapolating this rather steep X-ray spectrum to the Ly~edge.

The results of the photoionization model predictions and the observed 
 emission line parameters are listed in Table~1.
For each emission line observed, the ratio of the observed
 (Column~2) and dereddened (Column~3; $A_V = 1\mag{.}4$) line fluxes to the 
 H$\beta$ flux are tabulated along with the integral dereddened
 line luminosity (Column~4), the photoionization model predicted line
 luminosity (Column~5), and the residual line luminosity after subtracting
 the photoionization model prediction (Column~6).
The best-fit stellar continuum model combined with the best-fit photoionization
 model (convolved with the instrumental response function) is shown  
 as the middle (dotted) curve in Figure~\ref{fig:ospect_mod}
(displaced by $1/2$ for clarity).

\subsection{The contribution from shocks} \label{s:shocks}

These pure photoionization models predict strong \oiii, He~{\sc i}, 
 and H Balmer line emission excited by the UV emission from the underlying
 star cluster
 but are unable to reproduce the observed strength of several key emission lines
 (see Table~1
and Figure~\ref{fig:ospect_mod}).
A significant excess is observed in 
 \sii\ $\lambda\lambda$6716, 6731,
 [O{\sc i}] $\lambda\lambda$6300, 6364 
and [N{\sc ii}] $\lambda\lambda$6548, 6583~\AA\ doublets.
We find the additional dereddened luminosity in the \sii\ doublet, for 
 example, is $L_{\rm{[SII]}}=1.72\times 10^{38}$~\ergl.
Further, the residual \oiii\ $\lambda(4959+5007)/\lambda 4363$ ratio is
 low, $\sim$20, compared to typical photoionized \hii\ regions.

The observed excesses in these low-excitation lines and the low 
 \oiii\ ratio are likely caused by  
 shock excitation in the nebula (cf. \citet{mappings}). 
We can identify three plausible sources for shock excitation: 
({\it i}) individual SNe in the stellar cluster, 
({\it ii}) the cumulative stellar wind in the cluster, and 
({\it iii}) the ULX source. 
Here we make only rough estimates of the possible contributions from these three 
sources; we lack sufficient knowledge of the physical conditions in the nebula
 (e.g., shock velocity, pressure, density, temperature, and pre-shock ionization)
 to apply a full shock-excitation model for the emission lines. 

The Starburst99 models predict the total supernova production rate
 in a cluster of age $t_c \sim 4.25$~Myr and mass $M \sim 10^5$~\msun\
 and a Salpeter IMF is about $6\times 10^{-5}$~yr$^{-1}$.  
Since SNRs are visible in emission lines for
 up to $10^5$ years \citep{Cioffi}, the effective
 number of SNRs contributing to the spectrum is at most $\sim 6$. 
Typical H$\alpha$ and \sii\
 luminosities from an individual SNR are 
 $L_{{\rm H}\alpha} \sim L_{\rm [SII]} \sim (3-6) \times 10^{36}$~\ergl\ 
 \citep{braunm31,MF_cat}.
The expected luminosity in these lines from 6 SNRs is then 
 $\sim (1.8-3.6) \times 10^{37}$~\ergl, or 5 to 10 times less than the 
 residual luminosities listed in Table~1. 

At the estimated age of 4.25~Myr, the Starburst99 model predicts equal contributions
from SNe and stellar winds to the mechanical luminosity of the cluster. If the 
conversion efficiencies from mechanical luminosity input to emission line radiation
are the same for SNe and stellar winds, winds will contribute a comparable 
amount to the shock-excited lines. The combination of SNe and stellar winds still leaves 
the \sii\ $\lambda\lambda$6717, 6731 
production about 2.5 to 5 times below that needed to explain the observed luminosity. 

The third plausible source for shock excitation  is the ULX.
The presence of such an object in the cluster may contribute to
the residual line emissions if mechanical energy from the X-ray source can be
 imparted to its surroundings in the form of an outflow. It has been postulated 
that some ULXs are microquasars \citep{Kingetal01}, whose radiation is
collimated along the jet axis, or supercritical accretion disks with outflows
like that in SS433 \citep{FaMe01}. 

In the case of microquasars, outflow in the 
 form of a leptonic jet could occur during the low state and during outbursts
 associated with transitions within the very-high/intermediate states
 (Fender, Pooley, \& Gallo 2004; Corbel \etal\ 2003).
However, utilizing the known correlation between X-ray and radio luminosities for
 black holes in the low-hard X-ray state (Corbel \etal\ 2003;
 Gallo, Fender, \& Pooley 2003) to estimate the mechanical power in the jet and 
assuming the \cxo\ observation of \cxou\ was obtained when the source 
 was near Eddington, then the power 
 in the jet is $L_J \lesssim  10^{38}$~\ergl\ even allowing for additional
 transient ejection events (Fender, Belloni, \& Gallo 2004).
The mechanical power imparted to the surrounding nebula in the form
 of outflow from microquasar-like jets from the ULX is then  
 far less than needed to explain the residual line emission.
[Note that the Galactic microquasars like 1E~1740.7-2942 and GRS~1758-258
 (Mirabel \etal\ 1993), 
for instance, have presistent lobe-like emission
 but are old stellar systems lacking associations with young energetic nebulae.]

In the case of SS433-like objects, outflow is in the form of persisent
 heavy jets that impart a large fraction of their accretion energy into their 
 surroundings in the form of mechanical energy.
The observed kinetic luminosity in the two jets of SS433
is $\sim 2 \times 10^{39}$~\ergl\ \citep{Fab04}. 
This is comparable to the predicted mechanical luminosity from stellar winds and SNe
 in the star cluster associated with \cxou, $3.5 \times 10^{39}$~\ergl,
 at its age of 4.25~Myr.

Likely all three sources of shock excitation contribute to the observed 
 line emission. The ULX can be an important, and even dominant, source of 
 radiative and mechanical feedback in the overall energetics of the emission region.

\section{Discussion} \label{s:discussion}

We have presented optical photometric and spectroscopic data that 
 help to describe the physical environment surrounding the ULX \cxou\ in \ngc.
\cxou\ is spatially-coincident (within astrometrical errors) with a bright,
young, $t_c=4.25\pm0.25$~Myr, star cluster of mass
$M = (1.1\pm0.2)\times 10^5$~\msun.
If the ULX is a part of this bright stellar cluster, 
 then the progenitor of the compact accretor must have been at least as massive as 40-50~\msun\
 in order to have already evolved through the supernova stage. 
The companion star is also likely massive and evolved in order to provide the 
 high mass accretion rate needed to power the ULX.

The emission line spectrum of the surrounding nebula can be 
 interpreted as a result of photoinization produced
 by the cluster UV spectrum, but an additional source of shock excitation
 is needed to understand the observed \sii, [O{\sc i}] and [N{\sc ii}] 
 line intensities. 
The mechanical input from SNe and stellar winds from the star cluster 
 can account for only
$\sim$20--40\% of the required shock excitation so some additional energy 
source is needed. The obvious candidate is the ULX itself.

Emission nebulae associated with nearby ULXs are 
apparently a common phenomenon.
Although no formal census has been undertaken, we find potential associations
 with bright, extended optical structures for 67 of the 154 ULX
 candidates tabulated in the ULX survey of Swartz \etal\ (2004). 
Although these associations are based mainly upon Digitized Sky Survey images
 and must await more conclusive analysis, many detailed studies 
 of ULXs with associated nebulae have appeared in
 the recent literature \citep{Pakul_Mir03,Leh_ea05,
kuntzm101, Pakul06, GPM1, GPM2,Abol_ea07}.

In particular, the nebula associated with 
 \cxou\ in \ngc\ can be compared with  the compact nebula MF\,16 associated with the ULX
NGC\,6946~X-1 \citep{MF16_BFS}. 
The MF\,16 nebula is an isolated object: It does not appear to be 
 contaminated by associated star
 clusters or nearby H\,II regions. 
The line luminosities of MF\,16 \citep{MF16_BFS,Abol_ea06} are 
 $L_{{\rm [SII]}} \approx L_{{\rm H\alpha}} \approx 2 \times 10^{38}$~\ergl, 
 $L_{{\rm H\beta}} \approx 6
\times 10^{37}$~\ergl; equal to  
the residual line luminosities in the \ngc\ nebula (Table~1). 
On the other hand, \oiii\ $\lambda\lambda 4959, 5007$ and \oi\ $\lambda 6300$ \AA\
lines in MF\,16 are 3.5 times brighter than those in the \ngc\ residuals. These two lines
are excited in different ways: by UV photoionization in the case of
\oiii\ $\lambda\lambda 4959, 5007$ and by shock excitation in the case of 
\oi\ $\lambda 6300$ \AA. 

Additional sources of mechanical energy in the form of more or less collimated winds
or jets is not unprecedented in ULX nebulae.
In the nebulae of ULXs NGC\,6946~X-1 and
Holmberg\,II~X-1 (another ``isolated'' ULX nebula), radial velocity gradients of $50-100$~km~s$^{-1}$ have been found 
\citep{Leh_ea05,Fab_ea06}. It was concluded that the nebulae have to be
powered by the central ULXs and that, therefore, the ULXs must also provide a source of gas collisional
excitation in the ULX-associated nebulae. 
It was found recently through optical spectroscopy of 8 ULX-associated nebulae
\citep{Abol_ea07} that in at least half of them shock excitation
dominates. We cannot say definitely about the kinematical properties of the gas in 
the \cxou\ counterpart -- we lack both spectral and spatial 
resolution for this -- but it is likely the ULX plays an important role here as well. 

\acknowledgements
The authors thank N.\,Borisov for help with the observations.
This work has been supported in part by Russian
RFBR grants NN\, 06-02-16865 and 07-02-00909
and by a joint RFBR/JSPS grant N\,05-02-12710.
Additional support was provided by NASA under grant NNG04GC86G issued 
through the Office of Space Science and by the 
Space Telescope Science Institute under the grant HST/AR-10954.

\clearpage

\begin{table} \label{tab:lines}
\begin{center}
\caption{Integrated Emission Line Fluxes and Luminosities}
\begin{tabular}{lccccc} 
\hline \hline 
line & $F / F(H_\beta)$ & $F / F(H_\beta)$ & $L$ & $L_{mod}$& $L_{res}$\\ \cline{4-6}
     &                  & (dereddened)     & \multicolumn{3}{c}{10$^{37}$ \ergl} \\ \hline
 H$\delta$                 &0.05$\pm$0.07    &0.211$\pm$0.012  &4.5$\pm$0.3    &3.8  & 0.7\\
 H$\gamma$                 &0.34$\pm$0.02    &0.444$\pm$0.016  &9.5$\pm$0.5    &6.8  & 2.7\\
 $[$O{\sc iii}$]$ $\lambda$4363 &0.03$\pm$0.01    &0.04$\pm$0.02    &0.87$\pm$0.6   &0.0007&0.87\\
 H$\beta$                  &1.000$\pm$0.05   &1.000$\pm$0.006  &21.45$\pm$0.19 &14.59 &6.86 \\
 $[$O{\sc iii}$]$ $\lambda4959$ &0.35$\pm$0.04    &0.292$\pm$0.008  &6.28$\pm$0.17  &2.32  & 3.96\\
 $[$O{\sc iii}$]$ $\lambda5007$ &1.15$\pm$0.05    &0.919$\pm$0.006  &19.72$\pm$0.19 &6.98  & 12.74\\
 $[$N{\sc i}$]$ $\lambda5200$   &0.08$\pm$0.04    &0.050$\pm$0.006  &1.08$\pm$0.17  &0.004 &1.08\\
 $[$N{\sc ii}$]$ $\lambda5755$  &0.02$\pm$0.03    &0.011$\pm$0.005  &0.24$\pm$0.11  &0.006 &0.24\\
 HeI $\lambda5876$   &0.13$\pm$0.04    &0.085$\pm$0.005  &1.81$\pm$0.14  &2.01  & -0.19\\
 $[$O{\sc i}$]$ $\lambda$6300$+$
 $[$S{\sc iii}$]$ $\lambda6310$ &0.15$\pm$0.04    &0.085$\pm$0.007  &1.83$\pm$0.15  &0.01  &1.81\\
 $[$O{\sc i}$]$ $\lambda6363$   &0.02$\pm$0.02    &0.008$\pm$0.003  &0.18$\pm$0.09  &0.004  &0.18\\
 $[$N{\sc ii}$]$ $\lambda6548$  &0.668$\pm$0.003  &0.349$\pm$0.002  &7.48$\pm$0.05  & 0.33 & 7.15\\
 H$\alpha$                 &6.375$\pm$0.009  &3.360$\pm$0.005  &72.09$\pm$0.16 &44.77  &27.32 \\
 $[$N{\sc ii}$]$ $\lambda6583$  &2.00$\pm$0.009   &1.046$\pm$0.005  &22.43$\pm$0.15 & 0.97  & 21.46\\
 HeI $\lambda6678$   &0.04$\pm$0.03    &0.023$\pm$0.004  &0.50$\pm$0.11  & 0.57  &-0.07\\
 $[$S{\sc ii}$]$ $\lambda6717$  &0.97$\pm$0.02    &0.493$\pm$0.011  &10.5$\pm$0.3   & 0.6  &10.0\\
 $[$S{\sc ii}$]$ $\lambda6731$  &0.71$\pm$0.02    &0.356$\pm$0.011  &7.6$\pm$0.3    & 0.4 & 7.2\\
\hline
\end{tabular}
\end{center}
\end{table}

\clearpage

\clearpage

\clearpage

\end{document}